\documentclass{aa}
\usepackage{graphicx}
\usepackage{times}
\def\degr{\hbox{$^\circ$}}
\def\arcmin{\hbox{$^\prime$}}
\def\arcsec{\hbox{$^{\prime\prime}$}}

\def\farcs{\hbox{$.\!\!^{\prime\prime}$}}

\def\rj{r_{\rm jet}}

\def\msun{{\rm M}_{\sun}}
\def\mdotjet{{\dot{M}}_{\rm j}}
\def\mdotdisk{{\dot{M}}_{\rm a}}

\begin{document}

\title{Magnetic interaction of jets and molecular clouds in NGC\,4258}

\author{Marita Krause\inst{1}
\and Christian Fendt\inst{2}
\and Nikolaus Neininger\inst{1,3,4}
}

\offprints{M. Krause (mkrause@mpifr-bonn.mpg.de)}

\institute{Max--Planck--Institut f\"ur Radioastronomie,
               Auf dem H\"ugel 69,
               D--53121 Bonn, Germany
\and
           Max--Planck--Institut f\"ur Astronomie,
               K\"onigstuhl 17,
               D--69117 Heidelberg, Germany
\and
           Radioastronomisches Institut der Universit\"at Bonn,
               Auf dem H\"ugel 71,
               D--53121 Bonn, Germany
\and       Institut de Radioastronomie Millim\'{e}trique,
               300, Rue de la Piscine,
	       F-38406 St. Martin d'H\`{e}res, France
}
\date{Received ......... / Accepted .......}
\titlerunning{CO in NGC\,4258}
\authorrunning{Krause, Fendt \& Neininger}
\abstract
{
\object{NGC\,4258} is a well known spiral galaxy with a peculiar
large scale jet flow detected in the radio and in H$\alpha$. Due to the
special geometry of the galaxy, the jets emerge from the nuclear region
through the galactic disk -- at least in the inner region.
}
{Also the distribution of molecular gas looks different from that in other
spiral galaxies:
\element[][12]{CO}(1--0) emission has only been detected in the center and
along the jets and only up to distances of about $50\arcsec$ (1.8\,kpc)
from the nucleus.
This concentration of CO along the jets is similar to what is expected as
fuel for jet-induced star formation in more distant objects.
The reason for the CO concentration along the inner jets in NGC~4258
was not understood and is the motivation for the observations presented here.
}
{Using the IRAM interferometer at Plateau de Bure, we mapped the
\element[][12]{CO}(1--0)
emission of the central part of NGC\,4258
along the nuclear jet direction in the inner 3\,kpc.
In order to get a properly positioned overlay with H$\alpha$ we observed
NGC\,4258 in H$\alpha$ at the Hoher List Observatory of the
University of Bonn.
}
{
We detected two parallel CO ridges along a position angle of
$-25\degr$ with a total length of about $80\arcsec$ (2.8\,kpc),
separated by a CO depleted funnel with a width of about $5\arcsec$
(175\,pc).
The H$\alpha$ emission is more extended and broader than the
CO emission with its maximum just in between the two CO ridges.
It seems to be mixed in location and in velocity with the CO
emission.
In CO we see a peculiar velocity distribution in the iso-velocity map
and p-v diagrams.
We discuss different scenarios for an interpretation and present a model
which can explain the observational results consistently.
}
{
We propose here that the concentration of CO along the ridges is
due to interaction of the rotating gas clouds with the jet's
magnetic field by ambipolar diffusion (ion-neutral drift).
This magnetic interaction is thought to increase the time
the molecular clouds reside near the jet thus leading to the
 quasi-static CO ridge.

}
\maketitle
\section{Introduction}
\label{sec:introduction}
The nearby (D\,=\,7.2\,Mpc) spiral galaxy NGC\,4258 is known for its
anomalous H$\alpha$ and non-thermal radio arms located in between the
normal spiral arms (Court\`{e}s \& Cruvellier~1961, van der Kruit et al.\,1972).
The anomalous arms are usually explained in terms of collimated nuclear
outflows or jets emerging from the nuclear region. In the central region,
there is a bending at 70\arcsec (2.5\,kpc) radial distance in a trailing
sense with respect to the galactic rotation. The interpretation in terms of
nuclear jets is consistent with the detection of strong nuclear maser emission
from an edge-on, east-west orientated nuclear accretion disk (Watson \& Wallin~1994,
Greenhill et al.\,1995) with a super massive nuclear engine (Miyoshi et al.\,1995).\\
Cecil et al.\ (1995a,b) presented a high resolution radio continuum map which
shows the central radio emission to be nearly perpendicular to
the nuclear disk (Table~\ref{table1}) as expected for jets.
The radio jet can even be followed to mas scale with VLBA observations 
(Herrnstein et al. 1997, 1998a, 2005).\\
The geometical parameters of the galactic and the nuclear disk and the jet features
of NGC\,4258 are summarized in Table~\ref{table1}.

As the kpc-jet axis lies nearly along the major axis of the disk, the jets have to pass
through the galactic disk, at least in the central part, and an interaction with
the interstellar medium (ISM) of the disk of NGC\,4258 is expected.\\
Cecil et al.\ (1992) identified the jets in the SE in the
inner 2\,kpc as a helical, braided structure of 3 intertwined plasma
streams. The emission line velocity field suggests that the gas moves
along the helices. Already these authors assumed an interaction of the jets with the ISM of the galaxy. \\
Molecular gas (\element[][12]{CO}(1--0)) in NGC\,4258 has first been observed with the Kitt Peak 12-m 
telescope (Adler \& Liszt~1989), with the Nobeyama 45-m telescope (Sofue et al.~1989), 
and with the Owens Valley three element interferometer (OVRO)(Martin et al.~1989).
While the Kitt Peak observations detected CO in the central 60\arcsec\, but lack angular resolution to show details, the spatial and intensity distribution of the Nobeyama observations turned out to be not reliable enough. With the OVRO Martin et al. (1989) resolved the central emission in two components lying on each side of the nucleus and outlining a narrow channel through the interstellar disk corresponding to the H$\alpha$ jets. Entrained and shocked gas of the ISM were proposed to be responsible for
the optical line emissions. X-ray observations of the jets are interpreted by Cecil
et al.\ (1995a,b) as due to hot, shocked gas from nearby molecular clouds that
have been entrained into the jet.\\
The OVRO data were further analyzed by Plante et al.~1991 who suggested that the CO
distribution confines the inner part of the H$\alpha$ jets and that the line-emitting
jet may even be deflected by a dense molecular cloud. Cecil et al. (2000) claimed to
have detected two radio hot spots $24\arcsec$ north and $49\arcsec$ south of the nucleus and infered that the jets may have precessed on a large cone through the disk plane. 
The interpretation of these two radio sources as beeing hot spots is, however,
still under debate as the northern one has been classified as SNR candidate by Hyman
et al. (2001). Krause et al. (2004) detected linear polarization in the northern one but not in the southern. \\
The first reliable single-dish observations with enough angular resolution to resolve the \element[][12]{CO}(1--0) emission of NGC\,4258 had been caried out with the 30-m IRAM telescope at Pico Veleta (Krause et al.\ 1990). Molecular gas in NGC\,4258 has only been detected in the center and along the H$\alpha$ jets, at distances up to 2\,kpc (Krause et al.\ 1990). It has not been found in other parts of the (central) galactic disk up to the detection limit of about 30\,mK. Observations by Cox \& Downes (1996) confirm this.\\
In this paper we present high resolution Plateau de Bure (PdB) interferometer measurements of the molecular gas that were combined with single-dish data of the 30-m IRAM telescope. Hence, we present the \element[][12]{CO}(1--0) observations with the highest angular resolution ($4 \farcs 6 \times 3 \farcs 3$) obtained so far from NGC\,4258 without lacking for the extended emission due to missing spacings from interferometer data alone. In the meantime, Helfer et al. (2003) published maps of the CO emission of NGC\,4258 as part of the BIMA Survey of 44 nearby galaxies with a an angular resolution of $6 \farcs 1 \times 5 \farcs4$. Their interferometer observations were combined with the 12-m NRAO telescope on Kitt Peak. Their maps agree well with our results. These data were recently analysed together with \element[][12]{CO}(2--1) interferometer observations taken with the Submillimeter Array (SMA) on Mauna Kea by Sawada-Satoh et al. (2007). They also interprete their data as indicating interaction of the molecular gas by expanding motions from the nuclear region.\\ 
\begin{table}
\caption[]{Geometrical parameters of NGC\,4258}
\label{table1}
\begin{tabular}{lll}
\hline\noalign{\smallskip}
{\it Galaxy} \\
\noalign{\smallskip}
~Morphological type    &SAB(s)bc                   &(1)\\
~Distance              &$7.2\pm 0.3$~Mpc           &(2)\\
~                      &(1\arcsec corresponds to 35~pc)\\
~Inclination           &$72\degr$                  &(3)\\
~Position angle (p.a.) \\
~of major axis       &$-30\degr$                 &(3)\\
~p.a. of H{\sc i} bar  &$10\degr \pm 3\degr$       &(3)\\
\noalign{\medskip}
{\it Nuclear disk} \\
\noalign{\smallskip}
~Inclination           &$83\degr \pm 4\degr$       &(4)\\
~Position angle        &$86\degr \pm 2\degr$       &(4)\\
\noalign{\medskip}
{\it Jet features} \\
\noalign{\smallskip}
~p.a. in radio continuum &$-3\degr \pm 1\degr$ \enskip for $d<24\arcsec$  &(5,6)\\
                         &$-43\degr \pm 1\degr$ \enskip for $d>24\arcsec$ &(6)\\
~p.a. in H$\alpha$, CO         &$-25\degr$                    &(7)\\
~$\beta$ in radio continuum &$60\degr$   \enskip for $d<24\arcsec$  &(7,8)\\
                            &$75\degr$   \enskip for $d>24\arcsec$  &(7,8)\\
~$\beta$ in H$\alpha$, CO         &$83\degr$                    &(7,8)\\

\noalign{\smallskip}
\hline\noalign{\medskip}
\end{tabular}

\noindent
{\small
(1) de Vaucouleurs et al.\ (1976)  \newline\noindent
(2) Herrnstein et al.\ (1999)      \newline\noindent
(3) van Albada (1980)              \newline\noindent
(4) Miyoshi et al.\ (1995)         \newline\noindent
(5) Cecil et al.\ (2000)           \newline\noindent
(6) Krause \& L\"ohr \ (2004)     \newline\noindent
(7) this paper                     \newline\noindent
(8) $\beta$ is the angle between the nuclear jet axis and the rotational axis of the galaxy as discussed in Sect.~\ref{sec:orientation} adopting the warped disk model by Herrnstein et al. (2005) with a central disk inclination of $74\degr$.
}
\end{table}
In Sect.~\ref{sec:observations} we will describe our
observations in \element[][12]{CO}(1--0) and in H$\alpha$ and present the
observational results in Sect.~\ref{sec:obs.results}. The discussion
follows in Sect.~\ref{sec:discussion} and Sect.~\ref{sec:summa}
contains the summary and conclusions.

\section{Observations}
\label{sec:observations}
We observed the central region of NGC\,4258 along the inner H$\alpha$
jets up to a radial distance of about 2\,kpc in \element[][12]{CO}(1--0) using
the Plateau de Bure (PdB) interferometer in its compact
configuration, i.e.\ with baselines ranging from 24\,m to 176\,m.
In order to account for the elongated shape of the
central region we set up a mosaic consisting of 5 fields that were
aligned with the CO ridge known from previous single-dish observations
(Krause et al.\ 1990). The central and reference position of our
observations is $\alpha_{2000} = 12^h18^m57^s55,\, \delta_{2000} =
47\degr18\arcmin14.2\arcsec$.  The spacing between the individual
pointings was half the primary beam width (FWHM). The positions
relative to the center of the 5 fields of the mosaic are ($0\arcsec,0\arcsec$),
($10\arcsec,-20\arcsec$), ($18\arcsec, -40\arcsec$), ($-10\arcsec,20\arcsec$), ($-18\arcsec,40\arcsec$). These
positions agree within $2\arcsec$ with the major axis of
NGC\,4258. Thus, we covered with our mosaic an area of $40\arcsec
\times 100\arcsec$, with a uniform sensitivity along the major axis of
the resulting oval. The five fields were pointed at sequentially, spending 4
minutes for each pointing; after 20 minutes the phase calibrator was
observed and the sequence started again.\\
The correlator backends
were set such as to cover the whole IF bandpass with a resolution of
2.5\,MHz (6.3\,${\rm km\,s^{-1}}$) plus two units that were set to
cover the central velocity range with a resolution of 0.625\,MHz.\\
The flux and bandpass calibration was determined by observing 3C273
and W3OH. We used 1044+719 as phase calibrator. The 5 fields were combined
in a mosaic and subsequently cleaned using the MAPPING procedure of the
GILDAS software package. The maps were cleaned with varied clean boxes for the
different channels according to their intensity distribution.\\
We corrected the data for `missing short spacings' by combining them with
single dish data obtained with the 30-m telescope at Pico Veleta by
Cox \& Downes (1996). For the combination we used the 'ssc method' as described
by Weiss et al. (2001) that works on the finally reduced (cleaned and corrected
for primary beam attenuation) interferometer cube. Both data sets were transformed into the uv plane using an FFT algorithm. The inner visibilities of the interferometer data were replaced with the single-dish values. The
angular resolution of the final maps is $4 \farcs 6 \times 3 \farcs
3$ with a position angle of $31\degr$.\\
We further observed NGC\,4258 in H$\alpha$ at the Hoher List
Observatory of the University of Bonn. The WWFPP focal reducer/CCD
camera at the 1-m telescope gives a field of view of $20'$
diameter at $0.8''$ pixel size. The seeing was about $2''$ and the
integration time in the 4\,nm wide H$\alpha$ filter 32 minutes. A
broadband red image was taken at the same time to subtract the
continuum emission. The large number of cataloged stars in the field ensures a proper relative positioning of the H$\alpha$ and the CO emission with an error of the order of $1''$.\\

\section{Observational results}
\label{sec:obs.results}
\subsection{The CO spectra and channel maps}
\label{sec:spectra}

We received spectra ranging from $-250$ to $+250$\,km/s with respect to the systemic velocity $v_{\rm lsr}$ of 457 km/s (van Albada, 1980). Peak intensities reach about 300~Jy/beam in the strongest spectra. Many of the spectra have more than one velocity component (see Sect.~\ref{sec:velocity distribution}).

\begin{figure*}[htb]
\begin{center}
\includegraphics[bb = 53 77 536 575,angle=270,width=16cm]{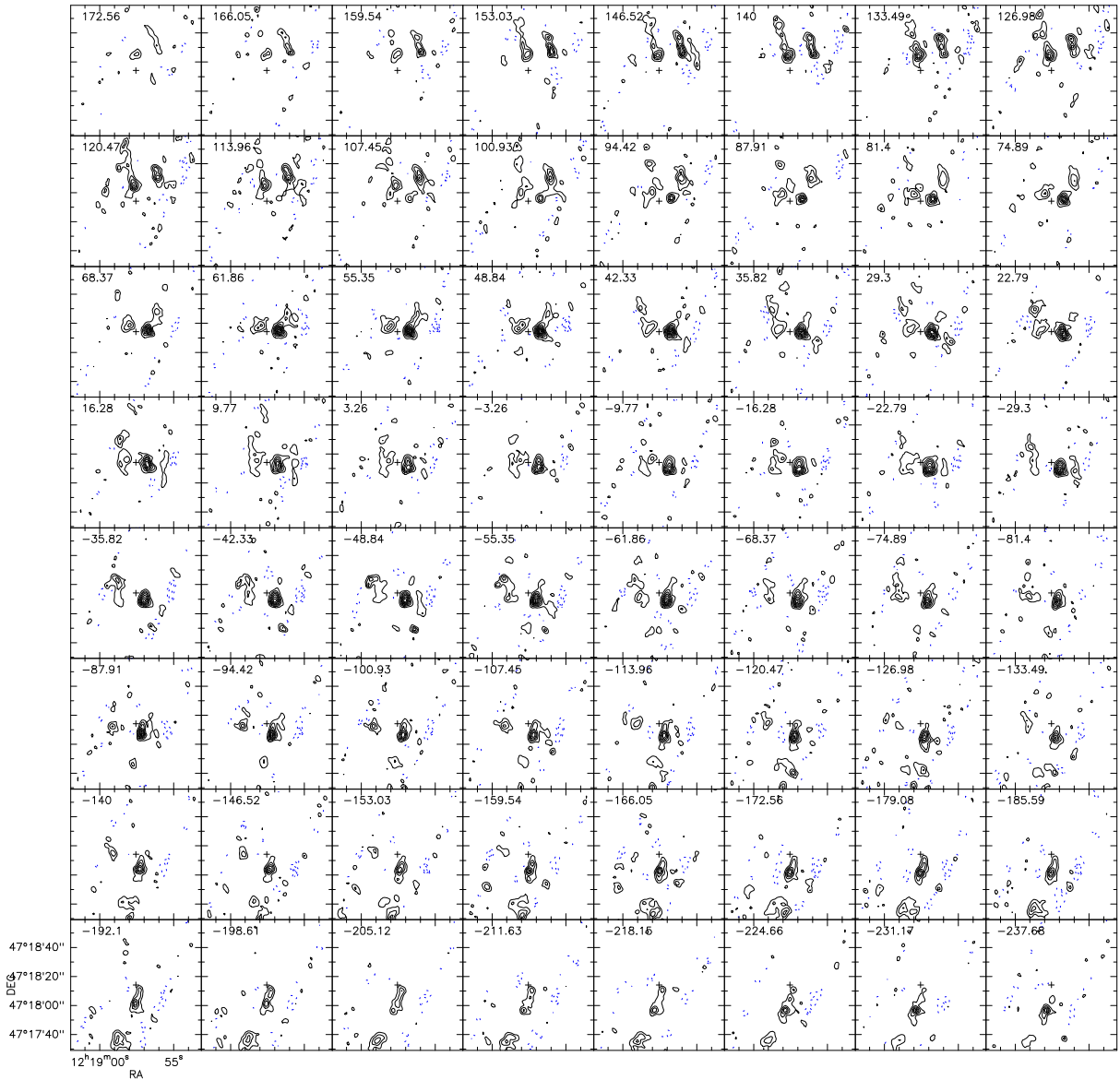}
\caption{Channel maps of \element[][12]{CO}(1--0) of NGC\,4258. The contour levels are 50mJy/beam. The velocity is given in the upper left corner in each panel.
\label{channel}
}
\end{center}
\end{figure*}

The central position of our observations (as given in Sect.~\ref{sec:observations}) agrees within less than 1\arcsec{} with the
position of the central maser emission as determined by Greenhill et
al.\ (1995). We could not detect CO right at the nucleus. In general,
there is a gap in the CO distribution along the central axis of the
inner H$\alpha$ jet. This is also obvious in the channel maps shown in
Fig.~\ref{channel}. In all panels in the velocity range between $-160$ to
$+160$~km/s show two main blobs visible oriented in east west direction
and separated by 15\arcsec{} to 20\arcsec{}.
The positions of the blobs shift together from north to south when changing
the velocities from positive to negative values.

\subsection{Integrated CO intensity and comparison with H$\alpha$}
\label{sec:intensity}
\begin{figure}[htb]
\begin{center}
\includegraphics[bb = 51 38 548 530,angle=270,width=6.5cm]{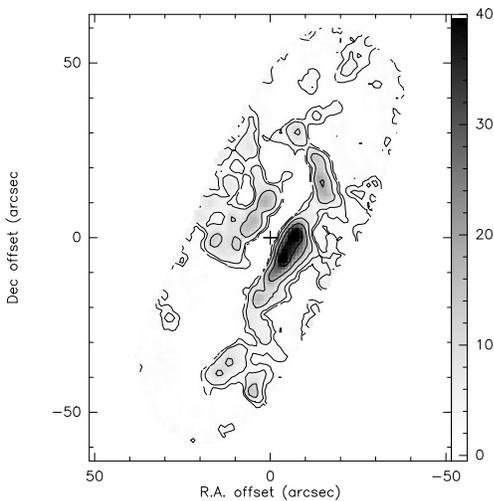}
\caption{Integrated intensity of \element[][12]{CO}(1--0) toward the central region of NGC\,4258. The rms-noise in the inner part of the map is 0.7\,Jy/beam km\,s$^{-1}$.
The contour levels are 2, 4, 10, 20, 35, 50 Jy/beam km\,$s^{-1}$. The cross marks the nucleus.
\label{mom0}
}
\end{center}
\end{figure}

We determined the zeroth moments by integrating the line intensity in the
individual spectra. All points with values below 3 times the noise level
($\sigma=0.7\,{\rm Jy/beam km\,s^{-1}}$) of the inner part of the map were
discarded in the calculation. The spatial distribution of the integrated CO emission in the observed area is shown in Fig.~\ref{mom0}.

\begin{figure}[htb]
\begin{center}
\includegraphics[bb = 57 43 529 490,angle=270,width=7.0cm]{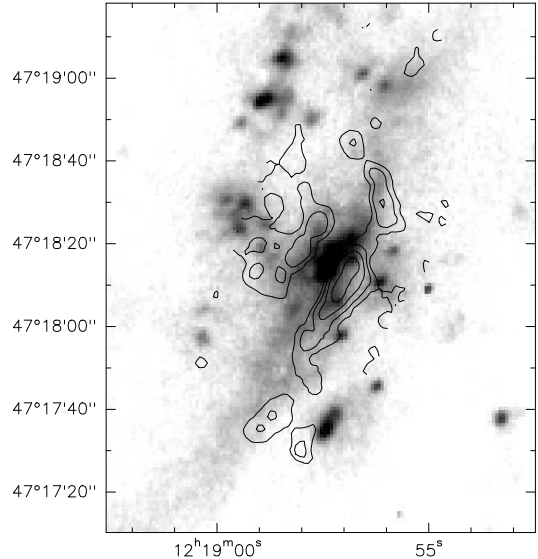}
\caption{Contours of the integrated intensity of \element[][12]{CO}(1--0) of
NGC\,4258 (equal to those in Fig.~\ref{mom0}) superimposed onto the
H$\alpha$ image of the galaxy. The normal spiral arms are located in
the north and south.
\label{COonHalpha}
}
\end{center}
\end{figure}

The molecular gas follows {\em two} parallel ridges along a position
angle of about $-25\degr$. The western ridge is stronger in intensity and
with a total projected length of about 80\arcsec{} more extended than the eastern one.  The width of each ridge (HPBW) is about
7\arcsec{} (250\,pc).  The ridges are separated by about 5\arcsec ~(175\,pc), a region with strong CO depletion.
The orientation of the ridges agrees well with previous
CO observations. (They had however either not enough spatial
resolution to separate both ridges (Krause et al.\ 1990; Cox \& Downes 1996)
or not enough sensitivity to trace both ridges (Plante et al.\ 1991, 1994).)  This
orientation matches the mean jet direction in the radio
continuum up to a distance of about $1.5\arcmin$ from the nucleus (see
Table~\ref{table1}).\\
Both northern ridges seem to kink by about $30\degr$ counterclockwise
at a distance of about $15\arcsec$ (530\,pc) from the
nucleus. At that position a kink in the western CO ridge has already
been observed by Plante et al.\ (1991) as component 2. It has been
interpreted as being due to a deflection of the jet by a dense
molecular cloud.\\
The increase of intensity in the outermost 10\arcsec{} of our observed
area is artificial and due to the increase of the rms-noise because of
the primary beam correction. This effect is most clearly visible along
the western edge of the map. \\
The superposition of the CO
distribution on our H$\alpha$ map is presented in
Fig.~\ref{COonHalpha} and shows that the CO ridges follow the
H$\alpha$ emission.

The H$\alpha$ emission is more extended than the CO emission. It has
its maximum in between the two CO ridges with a slight shift towards the
stronger western ridge. The H$\alpha$
emission however is broader than the CO emission and outlines the jet to
larger distances from the nucleus.  Contrary to previous
interpretations of Martin et al.\ (1989) and Plante et al.\ (1991, 1994) the CO
does not outline 'walls' around the H$\alpha$ jets and hence does not confine
it. It rather seems to be mixed in location and in velocity with the
H$\alpha$ gas.\\

\subsection{The velocity distribution}
\label{sec:velocity distribution}
\subsubsection{Iso-velocity and velocity dispersion maps}
\label{sec:iso-velocity}

\begin{figure}[htb]
\begin{center}
\includegraphics[bb = 146 38 548 479,angle=270,width=7.5cm]{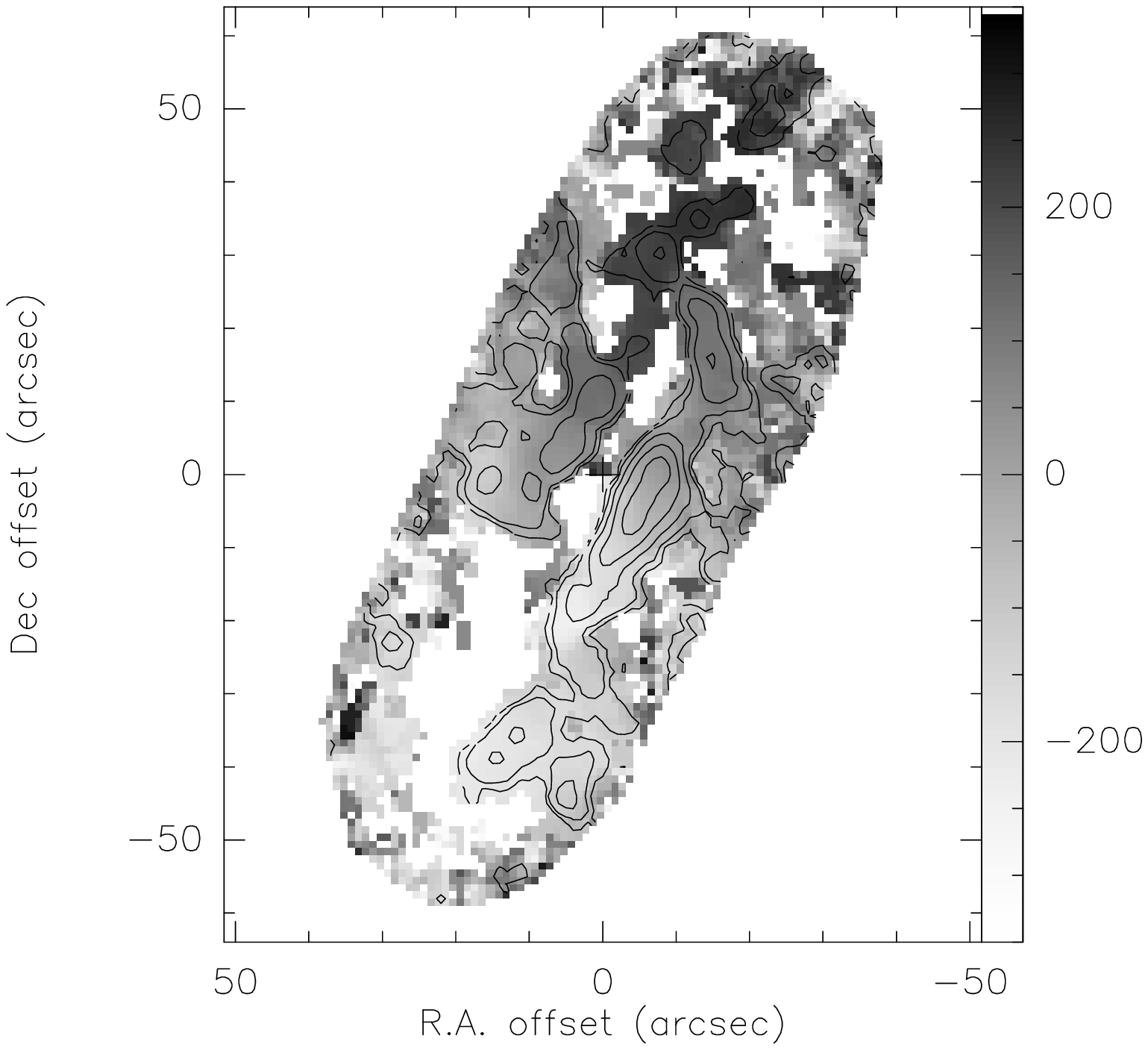}
\caption{Iso-velocity map of \element[][12]{CO}(1--0) with respect to
the systemic velocity in [km/s]toward the central region of NGC~4258 (grey scales). All points for which the integrated intensity is below its $3 \sigma$ noise level are discarded. The contours represent the integrated intensity
and are equal to those in Fig.~\ref{mom0}. The cross marks the nucleus.
\label{mom1+0}
}
\end{center}
\end{figure}

Fig.~\ref{mom1+0} shows the iso-velocity or first moment map of our
spectra. The velocities in the southern part of NGC\,4258 are blue shifted whereas those in the northern part are red shifted with velocities up to about $200~{\rm km\,s^{-1}}$ (north), resp. up to $ -200~{\rm km\,s^{-1}}$ (south).
This is indeed expected for normal galactic rotation and indicates that most of the molecular gas takes part in the galactic rotation.
A comparison with the rotation curves found in \ion{H}{i} and H$\alpha$ is given in Sect.~\ref{sec:pv_major}.

\begin{figure}[htb]
\begin{center}
\includegraphics[bb = 51 38 548 498,angle=270,width=6.0cm]{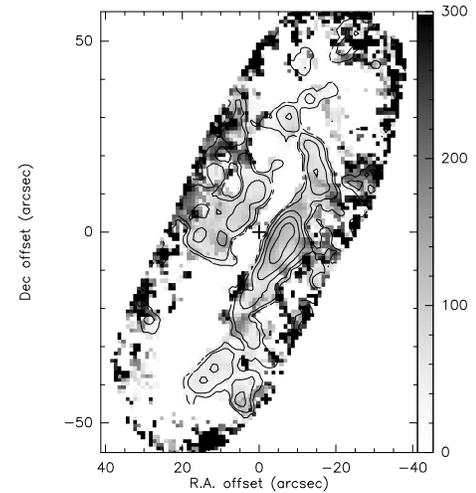}
\caption{Map of the velocity dispersion in [km/s] of \element[][12]{CO}(1--0) toward the central region of NGC\,4258 (grey scales).
Points for which the integrated intensity is below the $3 \sigma$
noise level have been omitted. The contours represent the integrated
intensity and are equal to those in Fig.~\ref{mom0}. The cross marks the nucleus.
\label{mom2+0}
}
\end{center}
\end{figure}

At close look the iso-velocity pattern reveals however a peculiarity: we observe a significant velocity gradient between the eastern and western side of the major axis of the galaxy at the same distance from the nucleus.
That is unusual for normal galactic rotation and will be discussed in Sect.~\ref{sec:scenario1}.\\
The velocity dispersion or second moment map is given in
Fig.~\ref{mom2+0}. For a better comparison we superimposed also here the
contours of the integrated intensity as presented in Fig.~\ref{mom0}.

It shows that the velocity dispersion along the ridges is generally
large with values of about 80 to 160~${\rm km\,s^{-1}}$ along the western
ridge up to a distance of about 25\arcsec{} from the nucleus. Further outwards
the velocity dispersion decreases to values of $\le 60 {\rm km\,s^{-1}}$.
The velocity dispersion seems to reflect the multiple component structure of many
spectra just in the inner part whereas the spectra farer away from the nucleus have only one component.

\subsubsection{p-v diagrams along the major axis of the galaxy}
\label{sec:pv_major}

\begin{figure*}[htb]
\begin{center}
\includegraphics[bb = 58 39 542 803,angle=270,width=16cm]{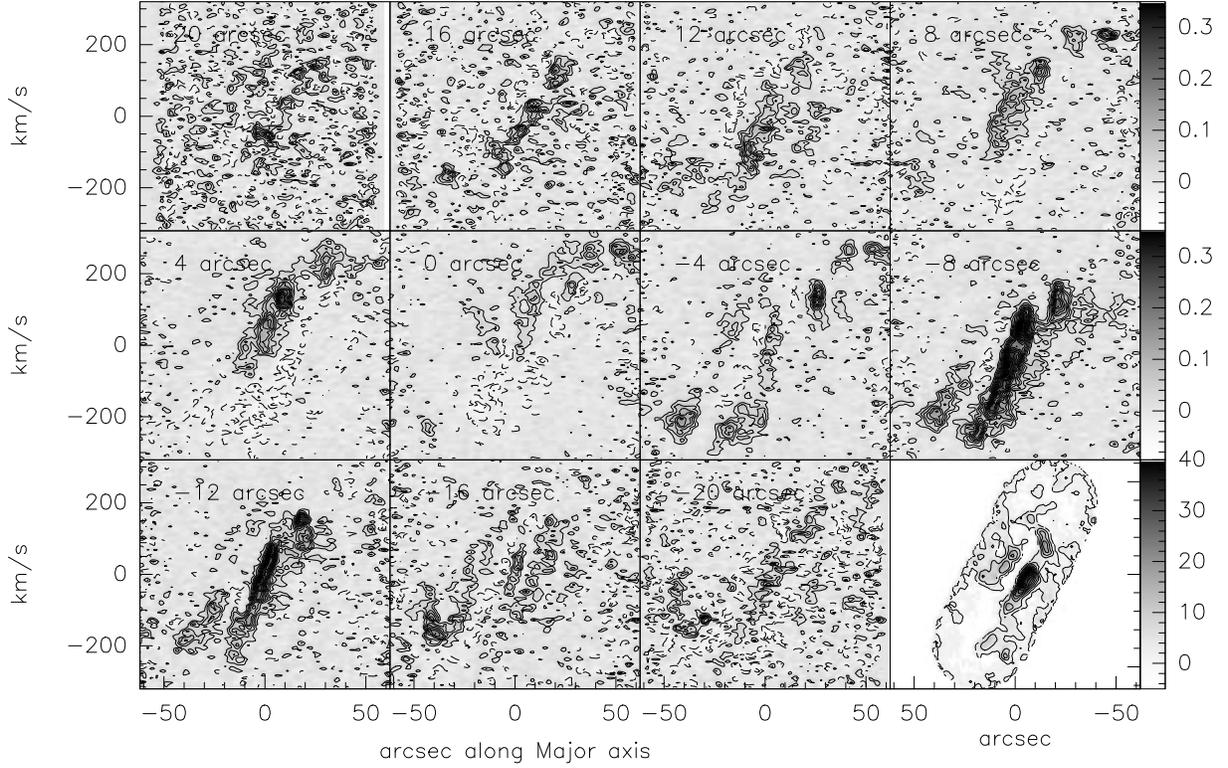}
\caption{Position-velocity diagram of NGC\,4258 along the CO-ridges and parallel
to it with distances from 20\arcsec east (upper left panel) to 20\arcsec west
(panel labeled with -20 arcsec).
The relative positions are as in Fig.~\ref{mom0}: -50\arcsec ~along the major axis
means southeast and 50\arcsec northwest of the nucleus.
The intensities are given in grey scale, their units are given by the wedges in
[Jy/beam km\,s$^{-1}$].
The lower right panel shows again the integrated intensity (as shown in Fig.~\ref{mom0})
for comparison.
\label{pv_major}
}
\end{center}
\end{figure*}

We determined several position velocity diagrams along the major axis of NGC\,4258
and parallel to it with a separation of 4\arcsec~to a distance of $\pm$ 20\arcsec
from the major axis.
The orientation of the major axis deviates only by 5\degr~from the orientation of
the CO ridges and was found to be a good linear approximation for the position
of the jet axis.
The principal features of the p-v diagrams are not sensitive to the exact position
of the axes chosen.
The p-v diagrams are shown in Fig.~\ref{pv_major} together with the integrated
intensity (as shown in Fig.~\ref{mom0}) in the lowest right panel.
As expected from the strong CO depletion between the two ridges in the integrated
intensity map, the p-v diagram along the major axis (position $0 \arcsec$) has
only very low intensity.
The highest intensities are found at the p-v diagrams at positions $ +4\arcsec$,
$ -8\arcsec$, and $ -12\arcsec$ east and west along the major axis.

The dominant shape and structure is elongated and rises from negative velocities
in the south-east to positive velocities in the north-west.
Such a feature is close to what is expected for CO gas rotating in a galactic disk.

A comparison of our p-v diagram with the H{\sc i} rotation curve by
van Albada (1980) shows that the CO velocities rise steeper than the
H{\sc i} velocities which reach velocities of about 200~${\rm km\,s^{-1}}$
only at distance of about 40\arcsec{} from the nucleus and then seem to
flatten further out.
The angular resolution of the H{\sc i} observations is however only 30\arcsec
along the major axis.
This means that we can only see an average of all H{\sc i} velocities between
e.g.\ 0\arcsec ~and 30\arcsec ~from the nucleus.
Hence the H{\sc i} velocities are only lower limits for the true rotation of the
ISM in NGC\,4258 and the deviation of our CO velocities
from the H{\sc i} rotation velocities does not contradict our interpretation of
CO being a rotating molecular disk gas. \\
Peculiar features in the pv-diagrams are the parallel structures that are visible
in some diagrams, mostly in those at $8\arcsec$ at positive velocities and at
$-4\arcsec, -8\arcsec, -12\arcsec$ at negative velocities.
It means that CO clouds with equal velocities are found at two different distances
from the nucleus along the corresponding cut.
This is also visible in the iso-velocity map in Fig.~\ref{mom1+0}.
The typical distance between the parallel structures in Fig.~\ref{pv_major} is about $20\arcsec$.

\subsubsection{p-v diagrams perpendicular to the major axis of the galaxy}
\label{sec:pv_minor}

\begin{figure*}[htb]
\begin{center}
\includegraphics[bb = 276 63 548 728,angle=270,width=14cm]{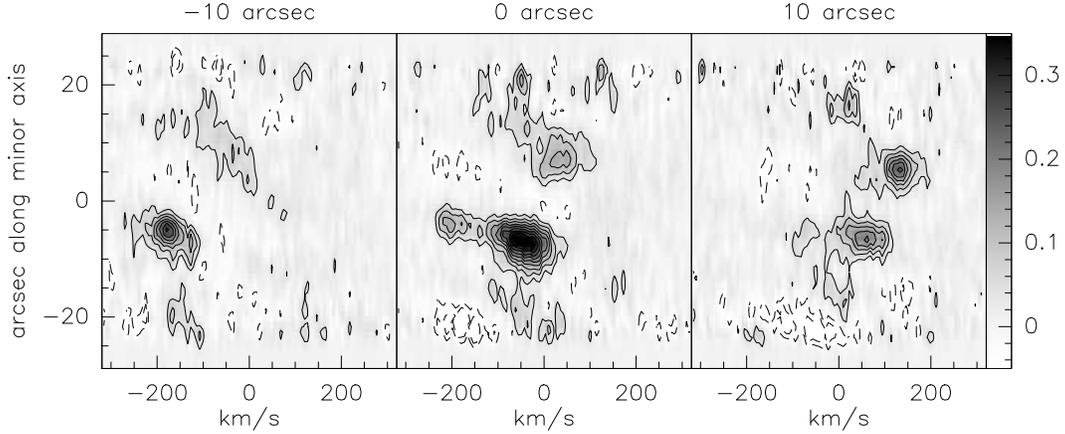}
\caption{Position-velocity diagrams of NGC\,4258 {\em perpendicular}
to the Co-ridges at distances $-10\arcsec$ (south), 0\arcsec, and
10\arcsec\ (north) from the nucleus. As in Fig.~\ref{mom0} 10\arcsec
~along the minor axis means northeast and $-10\arcsec$ southwest of the
nucleus.
\label{pv_minor}
}
\end{center}
\end{figure*}

We further determined p-v diagrams {\em perpendicular} to the major
axis (the jet axis), i.e.\ along the minor axis and parallel to it with
a separation of 5\arcsec~up to a distance of 40\arcsec~to the north
and south, respectively. Three characteristic ones are presented in Fig.~\ref{pv_minor}.

Up to a distance of 30\arcsec~from the nucleus most of these p-v diagrams perpendicular to the jet axis show two CO blobs shifted to
each other by about 15\arcsec~in position and about
50~${\rm km\,s^{-1}}$ in velocity.
These two blobs (we mean the two strongest
components in Fig.~\ref{pv_minor}) move together systematically
from negative (blue shifted) velocity to positive (red shifted)
velocity when one compares the different p-v diagrams from south to
north. The eastern blobs are generally more red shifted than the
western ones. A close comparison with the p-v diagrams {\em along} the
jet axis reveals that the blobs of CO emission show the systematic velocity shift between the 2 CO-ridges as described in Sect.~\ref{sec:iso-velocity}.

\subsection{CO mass estimate and distribution of the CO emission}
\label{sec:CO_mass}

The integrated CO flux in a 600\,km\,s$^{-1}$ wide band within the
area of emission (see Fig.~\ref{mom0}) is 630\,Jy\,km\,s$^{-1}$, respectively
102\,K\,km\,s$^{-1}$ (the data are corrected for missing spacings and primary beam
effects). This corresponds to a CO luminosity of
$2.6\times10^{8}$ K\,km\,s$^{-1}$pc$^{2}$
in agreement with the value found by Krause et al.\ (1990) and Cox \& Downes (1996)
with the Pico Veleta telescope (note however that the assumed distance to NGC\,4258 is slightly different in these papers).
Assuming the ``standard'' $N(\rm {H_2)}/I_{\rm CO}$ conversion
factor of $2.3\times 10^{20}$ mol\,cm$^{-2}$\,(K\,km/s)$^{-1}$
(Strong et al.\ 1988)
we derive a mass of molecular hydrogen of $M_{\rm H_2} \simeq 10^9~M_{\sun}$.
As already argued in Krause et al.\ (1990) this
mass derived from the conventional CO/$\rm H_2$ conversion may be an
unreliable estimate of the total mass of molecular hydrogen due to
possible extreme physical conditions along the jets. It is well in the
range of what is expected as {\em total} molecular hydrogen content in
the inner disk of a spiral galaxy, also those with a Seyfert~2 active
nucleus (e.g.\ Young \& Scoville 1991, Elmouttie et al.\ 1998).

In NGC\,4258 however, the CO emission is only concentrated along the inner
jets up to a radial distance of about 2\,kpc (Krause et al.\ 1990). No CO emission further outwards than 2\,kpc along the jets could be detected. This was confirmed by further observations of J.A. Garc\'{\i}a-Barreto (private communication) who extended the Pico Veleta CO map to the north up to a nuclear distance of 120\arcsec (3.8\,kpc)). He could also not detect CO emission along the northern normal spiral arm in NGC~4258.

The two blobs of CO emission north of the nucleus at a distance of about 50\arcsec that are visible in the CO map of the BIMA survey (Regan et al.\ 2001) are also visible in the Pico Veleta single dish map (Krause et al.\ 1990) but not within the area that was covered by our present PdB interferometer map. From the Pico Veleta map alone it was not really clear whether they belong to the northern spiral arm or the northern jet. The BIMA survey shows their (projected) position in between the normal spiral arm and the northern jet, their velocities (shown in Helfer et al.\ 2003), however, deviate strongly from those expected for the normal spiral arm at this nuclear distance. Hence we conclude that it cannot be regarded as a CO detection in the normal spiral arm in NGC~4258.\\

\section{Discussion}
\label{sec:discussion}

\subsection{CO as jet tracer}
\label{sec:tracer}
The main observed CO features, i.e. the concentration of CO in two
ridges along the jet axis as well as the lack of CO in other parts of
the galaxy may possibly be explained by an interaction process of the
{\em jet} with the interstellar medium.\\
Cox \& Downes (1996) proposed that the elongated CO feature traces
a {\em bar} flow in NGC\,4258 rather than being associated with the
nuclear jet.
An H{\sc i} bar has indeed been found in this galaxy (van Albada 1980)
but at a position angle deviating by $35\degr$ from the observed CO
structure (see Table~\ref{table1}).
Our high resolution observation now reveals, that
the CO has a funnel in between two parallel ridges.
The appearance of our p-v diagrams (Fig.~\ref{pv_major})
has not the rhomboid structure typical for bars
(e.g. Garc\'ia-Burillo \& Gu\'elin 1995).
Hence we exclude the bar hypothesis (which has already been rejected for other reasons by Hyman et al. ((2001) and Sawada-Satoh et al. (2007)) and assume that the CO structure is related to the jets.\\
The high amount of molecular gas observed along the jets and the lack
of CO elsewhere in the disk of NGC\,4258 already suggest that the
molecular gas was not ejected from the nucleus itself but originates from
galactic disk gas which has interacted with the jets.
This implies that the kpc-scale jet must be located more or less within
the galactic disk (see also Daigle \& Roy 2001).
The typical scale-height of the CO layer in galactic disks constrains the
deviation of the kpc-scale jet axis from the galactic plane to angles
between $1\,.\!\!\degr 5$ and $3\degr $ (Krause et al. 1990).

\subsection{The jet orientation}
\label{sec:orientation}
We now discuss observational constraints on the geometry of the jet
orientation.

The jet itself as seen in the radio emission shows a kink of
$40\degr $ at a radial distance of $24\arcsec$ (850pc)
symmetrically on both sides of the center (see Table~\ref{table1}).
The position angle of the H$\alpha$ and the CO feature is just the
mean jet direction in the radio continuum before and behind this
kink. The H$\alpha$ emission has always been interpreted as being related
to the jet (e.g. Cecil et al. 1992).

The position angle (p.a.) of the radio jet before the kink ($r<24\arcsec$)
is along the present spin axis of the accretion disk. Also a VLBA radio jet on sub-pc-scale has been
observed along this direction (Herrnstein et al. 1998a, Cecil et al. 2000).
Pringle et al. (1999), determined the angle $\beta$ between the {\em nuclear}
jet axis and the rotational axis of the galaxy
applying
\begin{equation}
\cos\beta = \sin\delta \sin i \sin\phi + \cos i \cos \phi.
\end{equation}
with the inclination of the galaxy $i$, the inclination of the inner
nuclear accretion disk $\phi$ (supposed to be perpendicular to the nuclear jet),
and the angle $\delta $ between the position angle of the major axis and that of
the projected nuclear jet direction (see Table~\ref{table1}).

Assuming $i=72\degr$, $\phi \simeq 82\degr$ (Herrnstein et al. 1999), 
and $\delta=27\degr$,  
we derive $\beta = 62\degr $ 
(similar to the values given by Pringle et al.~1999 or Wilson et al.~2001).
Minor deviations from these values exist in the literature.
For example, Wilson et al. (2001) discuss a disk inclination of $i=64\degr$.
A recent paper by Herrnstein et al. (2005) apply model fits of the 
maser data to a {\em warped disk} model, indicating warps in inclination 
as well as in position angle. 
The inclination warp of the inner accretion disk derived there follows a 
$i= 0.28 + 0.34 z$ distribution where $i$ is measured in radians and
$z$ in milliarcseconds along the l.o.s.
This results in a disk inclination of $\phi \simeq 74\degr$ close to
the center (i.e. for $z = 0$; see caption Fig.~10 or Eq.(14) for $r_i = 0$ in Herrnstein et al. 2005\footnote{Note that the authors plot the disk warping across 
the very center of the disk, however, the jet direction as indicated 
is not perpendicular to the disk.}).
Applying Eq. (1) for such a value will result in an offset angle
$\beta = 60\degr $.
The difference to the value derived above is negligible in spite 
of the rather large offset in the nuclear jet direction of about 
$ 8\degr = (82 - 74)\degr $.

In addition to the $60\degr$-offset in alignment between jet axis 
and rotation axis of the galaxy, it is worthwhile to note that 
the spin vectors of nuclear disk and galactic disk are actually
antiparallel, i.e. the offset between the spin vectors is $120\degr$
and not $60\degr$. This can easily be seen by comparing the Keplerian
rotation of the nuclear disk (e.g. Herrnstein et al.~1999) with e.g.
our CO data (Fig.~4).

As the jet inclination angle is close to $90\degr$ the knowledge
of the exact jet orientation would be essential if we would derive
e.g. the velocity structure of the nuclear jet due to the 
$1/cos(i)$-effect.
However, we do no calculate and do not need to know the velocity
of the nuclear jet, actually there are no velocity data for the
nuclear jet available.
In the next section we will discuss velosities of the kpc-jet and compare
them with our model scenario. This is independent of the nuclear jet 
orientation.

For the outer radio jet with $\delta=13\degr$ we derive $\beta = 75\degr$ ($73\degr$) for $\phi=82\degr$ ($74\degr$).
For the position angle of the H$\alpha$ jet (and the CO feature) we get
$\delta=5\degr$ and an angle between the H$\alpha$ jet and the
rotational axis of the galaxy $\beta = 83\degr$($80\degr$) for $\phi=82\degr$ ($74\degr$). 
(The values for $\beta$ with $\phi=74\degr$ i.e. the warped disk model (Herrnstein et al. 2005) are also summarized in Table~\ref{table1}.)
This is in rough agreement with the conclusion made above that the
H$\alpha$ jet and the CO feature must lie within the galactic disk. Again,
we point out that the p.a. of the radio jets before and after the kink
are just smaller and larger than the p.a. of the H$\alpha$ jet and the
CO feature.

The cause for such an apparent change in the direction of jet propagation is,
however, unclear.
We note that kinks and apparent changes in the jet flow direction
are common among extragalactic jets (e.g. Parma et al.\,1987).
Also, the NGC\,4258 jet tracers CO and H$\alpha$ show clear evidence for
inhomogeneities of the interstellar medium and deflections of the straight
jet propagation.

However, note that the change of direction between inner and outer radio
jet is {\em symmetric} for the jet and counter jet indicating on a systematic
mechanism working symmetrically on both sides of the jet.
Precession of the jet may be one of the possible mechanisms.
There are two indicators for ongoing precession observed.
One is the fact that the nuclear accretion disk is inclined towards the
equatorial plane of the galaxy (Miyoshi et al. 1995).
The other is the observed warping of the nuclear accretion disk (Herrnstein et
al. 1996, 2005).\\
If precession is really in action, the nuclear jet would precede along a cone
intersecting the equatorial plane of the galaxy at a certain time. It would
interact with the galactic disk just during the time of intersection and the
features of interaction (like H$\alpha$ and CO) would appear at those position
angles that belong to $\beta \simeq 90\degr $, which gives a
p.a.~$\simeq -28\degr$ (from Eq.\,(1)) of the observed features.
This is within the errors similar to the observed p.a. of the H$\alpha$ and CO features.
Therefore, the funnel observed in CO may also be caused by the jet propagation
during earlier times and probably with different strength.

For the following we will assume that the H$\alpha$ and CO features are related to
the jets in NGC~4258 and we will refer to their direction (p.a.) as 'kpc-scale jet' direction.
If we further assume that the kpc-scale jet lies within the galactic disk
(i.e. $\beta \simeq 90\degr$), the inclination $i_{\rm jet}$ of the jet towards the line
of sight is
\begin{equation}
\tan i_{\rm jet} = (\sin\delta\,\tan i)^{-1}.
\end{equation}
With $\delta = 5\degr$ and $i = 72\degr$ it follows $i_{\rm jet} = 75 \degr$. This means that the northern jet points towards the observer (and the southern jet away from the observer) as its position angle is west of the northern major axis of NGC\,4258.
However, a jet inclination of $75\degr$ also means that the jet lies close to the plane
of the sky. As we observe strong kinks in the jet at $d = \pm 24\arcsec$ in the position angle we also have to include the possibility of kinks in the direction along the line of sight. Such a kink (which is not detectable in our line observations and maybe also hidden in the $H\alpha$ observations) could lead to a (projected) geometry where the jets may be pointing away from us in the northern half (and towards us in the southern half).
The two different geometric models lead, however, to different interpretations
of the observed velocities in the line observations and will be discussed in the next
Section.

\subsection{Jet propagation and CO ridges}
\label{sec:orientation2}

We will now discuss both geometric scenarios and their implications on the question
how molecular gas can have been accumulated along the kpc-scale jet funnel.

In the first scenario (i) the angle between the jet axis and the
plane of the sky varies from a small positive (negative) value to a
small negative (positive) value. The northern jet axis points away from the observer.
The observed CO represents molecular gas entrained by the jet.
The jet accelerates the molecular gas observed in CO leading to a linearly
increasing velocity.
This kind of scenario has its similarities in molecular outflows driven by
protostellar jets. As it will be discussed in Sect.~\ref{sec:scenario1} this
scenario has to be rejected as an interpretation of the observed CO distribution
and velocities.

In the second scenario (ii) the jet inclination {\em along the jet}
remains constant and is along the galactic disk of NGC\,4258.
The northern jet axis points towards the observer.
The observed CO represents molecular gas in galactic
rotation. The concentration of CO just along the jet ridges is interpreted
as accumulation due to interaction of the rotating gas clouds with the
jet's magnetic field.
This scenario will be discussed in
Sect.~\ref{sec:scenario2} to Sect.~\ref{sec:time scales} and seems to be able
to put the puzzle of observational results into one consistent picture.

\subsubsection{Scenario (i): A jet-driven molecular CO-flow?}
\label{sec:scenario1}
Observations of protostellar outflows showed that large-scale mass
entrainment and acceleration of this mass by a
high-velocity jet may lead to a linear increase in the velocity
of the molecular outflow (Stahler 1994, Lada \& Fich 1996).
Time-dependent magnetohydrodynamic simulations of jet
formation (e.g.\ Ouyed \& Pudritz 1997, Fendt \& Cemeljic 2002)
found a similar increase of jet velocity with distance from the
jet source.
In our case, in NGC\,4258 the velocity distribution observed
along the CO ridges shows as well a linear increase with distance
from the center.
If we interpret these data in analogy to the protostellar
outflows, the linear increase seen in the CO velocity would
correspond to that of a continuously accelerated
jet-driven molecular flow.
We denote this model scenario with (i) and discuss its validity
in the following.

We first estimate the kinematic properties of the jet and
the potential molecular flow.
The jet mass flow rate is not known from observations.
In general, observations of astrophysical jets as well as theoretical
arguments (e.g.\ Ferreira 1997) showed that typically 1\% -- 10\%
of the disk accretion rate becomes ejected into the jets.
In the case of NGC\,4258, various number values for the
nuclear accretion rate were published so far.
Gammie et al. (1999) obtained
$\mdotdisk \geq 1.5\times 10^{-4} \msun {\rm yr}^{-1}$
for a thin warped disk, and
$\mdotdisk \simeq 10^{-2} \msun {\rm yr}^{-1}$
for an advection-dominated disk.
A standard accretion disk model gives
$\mdotdisk \simeq 7 \times10^{-5} \alpha \msun {\rm yr}^{-1}$
(Neufeld \& Maloney 1995)
with a standard accretion disk viscosity parameter $\alpha$.
Both alternatives, the advection dominated disk and the standard
accretion disk model seem to be consistent with the observations
(Herrnstein et al.\,1998b).
Yuan et al. (2002) explained certain spectral properties by shock
emission from the jet basis, thus lowering the nuclear jet mass
loss rate to about
$\mdotjet \simeq 3\times10^{-4} \msun {\rm yr}^{-1}$.
Recent observations by Modjaz et al. (2005) revealed similarly
low accretion rates indicating an upper limit of
$\mdotjet \simeq 10^{-3.7} \msun {\rm yr}^{-1}$.

In the following we adopt a nuclear accretion rate of
$\mdotdisk \leq 10^{-4} \msun {\rm yr}^{-1}$ corresponding to
an upper limit for the nuclear jet mass flow rate $\mdotjet \leq 10^{-5} \msun {\rm yr}^{-1}$.
With that we estimate the kinetic power of the {\em nuclear jet}
(i.e. the jet nearby to the accretion disk) as

\begin{equation}
L_{\rm j} \simeq 6\times10^{39}{\rm erg\,s^{-1}}
\left(\frac{v_{\rm jet}}{0.1c}\right)^{\!\!2}
\!\left(\!\frac{\mdotjet}{0.1\,\mdotdisk}\!\right)
\!\left(\!\frac{\mdotdisk}{10^{-4}\,\msun{\rm yr^{-1}}}\!\right)\!\!.
\end{equation}
This value is uncertain as both the nuclear jet velocity
and mass loss rate in NGC\,4258 are unknown.
Cecil et al.\ (1995b) discussed a nuclear jet velocity of $0.1\,c$.
However, there are frequent cases where nuclear jets in AGN propagate
with velocities close to the speed of light.
With a nuclear jet velocity of $0.9\,c$ the kinetic luminosity
increases to about
$L_{\rm j} \simeq 5\times10^{41}{\rm erg/s}$,
a value which is similar to the nuclear jet radio power derived by Falcke \& Biermann (1999).
Clearly, if the nuclear jet kinetic power would be entirely
tranformed into radio emission,
no energy available to drive the kpc-scale jet seen in
H$\alpha $ or CO.
In this respect,
our estimate for the kinetic power represents a lower limit.

An essential property for our discussion below is the
{\em kinematic time} scale of the kpc-scale jet
(i.e. the jet at distance of the order of kpc) which has to
be compared to the typical time scale of other processes
possibly involved.
The kinematic time scale can be derived
from the jet characteristic length and velocity, but
(as already indicated) the jet speed is not really known.
Several approaches to estimate the jet velocity have been
discussed in the literature.
Cecil et al. (1992) considered H$\alpha $ data and derived
a ballistic jet velocity about 2000\,km/s for a model of
"braided" jet "strands" ejected from orbiting nuclear
sources.
However, such a scenario seems to be ruled out by the
observations of the nuclear jet and the surrounding
masering disk.
The hot ionized gas observed in the X-ray band indicates
shock heating with inferred shock velocities of about
350 - 500\,km/s (Cecil et al. 1995a,b).
In general, the actual jet velocity depends on the physical
(hydrodynamic) conditions of the shock environment
and is typically several times larger.
Daigle \& Roy (2001) investigated possible jet trajectories
across the galactic disk discussing a possible velocity range
between 500 and $10^4$\,km/s.
They finally applied 4000\,km/s as kpc-scale jet velocity in NGC~4258.
Eventually, we estimate the kinematic time scale of the kpc-scale jet
by considering the extension of the H$\alpha $ emission
of $l_{{\rm H}\alpha} \simeq  12$\,kpc (see Cecil et al. 2000)
and a jet velocity derived from the H$\alpha $ emission
$v_{{\rm H}\alpha } \simeq 2000\,$km/s as a representative
value discussed in the literature,
\begin{equation}
\tau_{\rm kin} \equiv \frac{l_{{\rm H}\alpha}}{v_{{\rm H}\alpha }} \simeq
6\times10^6{\rm yrs}
\left(\frac{12\,{\rm kpc}}{l_{{\rm H}\alpha}}\right)
\left(\frac{2000\,{\rm km\,s^{-1}}}{v_{{\rm H}\alpha}}\right)^{-1}.
\end{equation}
Note, that the fact that the 'anomalous arms' (radio) in NGC\,4258 have
an extent larger than the whole galaxy (Hummel et al.\ 1989) may indicate
that the nuclear 'jet activity' has been present for a considerably longer
period, and the corresponding kinematic time scale and the amount of
mass injected into the jet from the nuclear accretion disk could
be appropriately larger.

Within the kinematic time scale of $6\times10^6{\rm yrs}$
a total mass of about $10^2\msun$ 
can be accumulated by the nuclear jet in the case of a mass loss
rate of $\mdotjet \simeq 10^{-5} \msun{\rm yr}^{-1}$.
This is in any case far below the value of the accelerated mass
inferred from the H$\alpha $ observations
($1.5\times 10^6 \msun$, Cecil et al.\ 1992) and CO observations
($10^9 \msun$, this paper).
Hence, one may conclude that most of the material observed in
H$\alpha $ and
CO cannot be launched at the jet source but could only
be {\em entrained} from the interstellar medium.

If we further follow the model assumption in this section and
interpret the CO velocity distribution observed in NGC~4258 as
linear velocity of a molecular flow assuming an inclination of
that flow of $i_{\rm jet} \simeq 75\degr$ as derived in Eq.\,(2),
the de-projected molecular flow velocity is
$(200\,{\rm km\,s^{-1}}/\cos75\degr)\simeq 800$\,km/s.
This would correspond to the velocity of the gas entrained
by the jet and would be in rough agreement (within a factor of two) with
the jet velocity estimated above (but is of course highly uncertain
due to the high inclination).
Therefore, from the interpretation of the velocity information
we have, the model of an entrained molecular flow would be feasible.

However, it would be difficult to understand how this amount of
molecular mass
could actually have been accelerated by a jet with parameters as estimated
for the NGC\,4258 nuclear jet
as there is obviously a strong mismatch between the kinetic momentum of both
components.
The kinetic momentum of the nuclear jet
$P_{\rm jet} \simeq 3\times10^2\msun \cdot 0.1\,c = 2\times 10^{45} {\rm g\,cm\,s^{-1}}$
is much lower compared to what is observed for the kpc-scale molecular flow
in CO or H$\alpha$ (assuming that the observed velocity distribution refers to
a linear motion).
Unless we assume much higher jet mass flow rates of the nuclear jet in the
past (probably triggered by a higher accretion rate) or a higher jet
velocity (e.g.\ $>0.9\,c$), we must dismiss
this scenario of a jet-driven molecular flow as proposed previously
in this section.
This holds even more as this scenario is unable to explain some other
observational results like the total CO intensity distribution and the
detailed structure in the p-v diagrams.

\subsubsection{Scenario (ii):
               A current carrying jet capturing interstellar CO}
\label{sec:scenario2}
If we consider model scenario (ii) to explain our observations,
the observed CO can be interpreted as molecular disk
gas in galactic rotation.
Its rotational velocity rises strongly in the central region up to a distance
of about 8\arcsec ~(280pc) and seemingly flattens further outwards.

The p-v curve along the major axis (see Fig.~\ref{pv_major}) is steeper
than the rotation curve observed in H{\sc i} (van Albada 1980) but not
as steep and irregular
as the velocities observed in H$\alpha$ by Dettmar \& Koribalski (1990)
and Cecil et al.\ (1992).
A steeper rise of the velocities in the central part of a galaxy in CO
compared to H{\sc i} has already been observed in many galaxies
(e.g.\ Sofue, 1997) and is rather expected due to the very different
angular resolutions in both maps as discussed in Sect.~\ref{sec:pv_major}.
Hence the CO in the disk of NGC\,4258 is to first order in 'normal' galactic rotation.

\begin{figure}[htb]
\begin{center}
\includegraphics[bb = 0 0 568 468,angle=0,width=8.8cm]{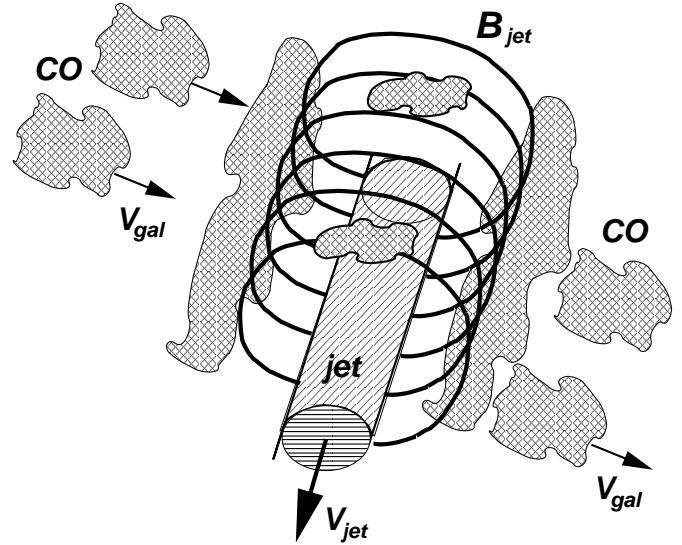}
\caption{Model scenario (ii) of the jet in NGC\,4258. The highly
collimated jet stream carries a net poloidal current equivalent
to a surrounding toroidal magnetic field. CO clouds in galactic rotation interact with the jet's toroidal magnetic field.
Most of the CO continues with the galactic rotation after being
slightly decelerated across the jet.
\label{model}
}
\end{center}
\end{figure}

However, the kinematics of the molecular gas are also affected by the kpc-scale jets.
The {\em non-detection} of CO outside the region along the inner jets
can be explained by an interaction of the molecular gas with the jets.
The CO clouds become temporarily `captured' leading to a longer staying time
of the gas near to the jets than elsewhere (see Fig.~\ref{model}).
We suggest that the reason for this interaction is a toroidal magnetic
field of the current-carrying jets (an interaction process being somewhat
similar to the formation of spiral arms due to a perturbation of the gravitational potential).

This jet-cloud interaction hypothesis is supported by the distinct
velocity and spatial structure of the observed CO which becomes obvious in
Fig.~\ref{pv_minor}. The gap in the CO intensity between the two blobs
in the p-v diagrams perpendicular to the major axis
can be interpreted as the central funnel
of the jets which is depleted of molecular gas and indeed follows the
gap between the two ridges in Fig.~\ref{mom0}.

The relative velocity shift between the two blobs is approximately
constant along the jet axis.
It is such that the CO velocities (absolute values) on the upstream side (with respect to the rotation of the disk gas) of the jets are smaller than those on the downstream side. This and also the double structure in Fig.~\ref{pv_major} clearly indicates
that interaction between the molecular disk gas and the jets is present and
{\em influences the rotation} of the molecular gas.

The observation of H$\alpha$ fits into our picture as signature of
former molecular gas which has been shock-ionized during the partly
violent interaction processes with the jets (as already proposed by Cecil et al.\ (1992).

Having given the motivation for our model hypothesis of a magnetic jet-cloud interaction in NGC\,4258, a more detailed discussion requires estimates of various physical parameters such as kinetic power, magnetic field strength, and
corresponding time scales.

\subsubsection{The jet magnetic field}
\label{sec:Bfield}
Polarimetric VLA observations of water vapor maser emission close to the
central engine of NGC\,4258 gave an upper limit for the parallel component
of the nuclear disk magnetic field (i.e. the toroidal field) at a disk radius
of 0.2\,pc of about 90 \,mG (Modjaz et al. 2005).
Note that is distance corresponds to about 50\,000 Schwarzschild radii from the
central black hole and is certainly not identical to the origin of the jet
which is expected to be as close to the black hole as about 10-50 Schwarzschild
radii.

Nevertheless,
theoretical models of magnetized jets suggest that the disk electric
current
$I_{\rm disk} \propto r_{\rm disk} B_{\phi ,{\rm disk}}$
is conserved.
We may therefore take the observed toroidal field strength at 0.2~pc and
extrapolate it to the inner (unresolved) disk region of jet formation
\begin{equation}
I_{\rm disk} \simeq {\rm const} = 3\,10^{17} {\rm A}
\left(\frac{r}{0.2\rm pc}\right)
\left(\frac{B_{\phi ,{\rm disk}}}{100\,{\rm mG}}\right).
\end{equation}
Assuming that the electric current in the disk-jet system is
roughly conserved,
$I_{\rm jet} \simeq I_{\rm disk}$,
we estimate the kpc-scale jet toroidal magnetic field as

\begin{eqnarray}
B_{\phi, \rm jet}
&\leq & 0.2\,{\rm mG}
\left(\frac{\rj}{100\,{\rm pc}}\right)^{-1}
\left(\frac{I_{\rm jet}}{3\times10^{17}{\rm A}}\right)\nonumber \\
& = & 0.2\,{\rm mG}
\left(\frac{r_{\rm disk}/\rj}{0.002}\right)
\left(\frac{B_{\phi ,{\rm disk}}}{100\,{\rm mG}}\right),
\end{eqnarray}
where here $\rj$ represents a measure for the kpc-scale jet radius.
This is consistent with accretion disk energy equipartition arguments
suggesting a poloidal current of about $10^{18}\,$A for active galactic
nuclei accretion disks (e.g. Camenzind 1990, Blandford et al.\,1990).

At larger distances from the nucleus radio observations detected a large
scale magnetic field along the anomalous arms
(van Albada \& van der Hulst 1982, Hummel et al.\,1989).
Recent radio continuum observations revealed a total magnetic field strength of
about $300\,\mu {\rm G}$ in the northern jet at about 5kpc distance from the
nucleus (with a linear resolution of about 100~pc) (Krause \& L\"ohr 2004).
This value is much higher than the usual field strengths measured in spiral galaxies
($1-50\,\mu {\rm G}$), but agrees surprisingly well with
the above estimate in Eq.\,(5).
Note that the above mentioned argument of electric current conservation
naturally links the observationally detected field strength in the nuclear
disk and the kpc-jet.
In practice, dissipative processes will be present in the jet and in the
jet-ambient medium interaction layer which will decrease the jet electric
current.

\subsubsection{Molecular clouds -- time scales and ambipolar diffusion}
\label{sec:time scales}

Molecular clouds are poorly ionized with a typical degree of ionization
of $10^{-7}$ (Pudritz 1990, Shu 1992), and {\em ambipolar diffusion}
plays an important role for the magnetic field interaction of such clouds.
Ambipolar diffusion considers the drift between the ionized and the
neutral component of the gas where both are coupled by dynamical
friction due to collisions and hence may affect the orbital motion of the molecular clouds if they cross the jet magnetic field. Such an interaction may decelerate the molecular clouds and therefore enhances the probability of finding them along the jet's edges, essentially forming a molecular gas-depleted jet funnel (see Sect.~\ref{sec:Inter}).

We now estimate the kinematic time scale for the jet-cloud {\em interaction}
from the travel time of a molecular cloud across the jet assuming
a width of the jet funnel
\footnote{
Here, $r_{\rm jet}$ is not associated with the actual radius of the
jet mass flow
but is a measure of the scale length (width) of the jet toroidal
magnetic field} of $2 r_{\rm jet}$.

The {\em dynamical} time scale for a magnetized molecular cloud is
given by $\tau_{\rm dyn} = l_{\rm cl}/v_{\rm alf}$,
where $ l_{\rm cl}$ is the typical size of the cloud
and $v_{\rm alf} = B/\sqrt{4\pi\rho}$ denotes the local Alfv\'en speed.
The time scale for {\em ambipolar diffusion} - the diffusion of the
magnetic field relative to the neutral particles under the influence
of collisions with ionized particles - is
$\tau_{\rm ad} = l_{\rm cl} /v_{\rm d}$
where
$v_{\rm d} = v_{\rm alf}^2 / (\sqrt{\rho}l_{\rm cl} C \gamma_{\rm ad})$
is the drift velocity with\footnote{
See e.g. Shu 1992 for numerical values of the parameter
$\gamma_{\rm ad} = 3.5\times 10^{13}{\rm cm}^3{\rm g}^{-1}\,{\rm s}^{-1}$
and
$C = 3\times 10^{-16}{\rm cm}^{-3/2}{\rm g}^{1/2}$
}
the drag coefficient $\gamma_{\rm ad}$ and a constant $C$.

Assuming that a jet magnetic field of strength of $200\,\mu {\rm G}$ (Eq.\,(5)) has penetrated the molecular cloud and considering molecular hydrogen
$\rho_{\rm cloud} = 2 n_{\rm cl} m_{\rm H}$, we obtain
\begin{eqnarray}
\frac{\tau_{\rm ad}}{\tau_{\rm dyn}} & =      &
\sqrt{4\pi} \frac{\gamma_{\rm ad} C l \rho}{B} \\ \nonumber
  & \simeq &
20 \left(\frac{ l_{\rm cl}}{10\,{\rm pc}}\right)
\left(\frac{ n_{\rm cl}}{10^3\,{\rm cm}^{-3}}\right)
\left(\frac{ B_{\rm jet}}{200\,\mu{\rm G}}\right)^{-1},
\end{eqnarray}
where $n_{\rm cl}$ is the hydrogen particle density.
This corresponds to a drift velocity of
\begin{equation}
v_{\rm d} = 0.5\,{\rm km\,s}^{-1}
\left(\frac{B_{\rm jet}}{200\,\mu{\rm G}}\right)^{2}
\left(\frac{n_{\rm cl}}{10^3\,{\rm cm}^{-3}}\right)^{-\frac{3}{2}}   
\left(\frac{l_{\rm cl}}{10\,{\rm pc}}\right)
\end{equation}
which is clearly below the mean galactic rotational velocity of the
molecular clouds.

In summary, our rough estimate indicates an order of magnitude difference in the time
scales $\tau_{\rm ad} \simeq 20 \tau_{\rm dyn} $ suggesting that the scenario of
an ambipolar jet magnetic field-cloud interaction is feasible.
If the dynamical time scale had been much larger than the ambipolar
diffusion time scale the (neutral) matter would hardly be affected by
the magnetic field. The other extreme case would correspond to a 'frozen-in' scenario of field and matter.

\subsection{Interaction of molecular clouds with the jet magnetic field}
\label{sec:Inter}
In the previous sections we have discussed important estimates of our
model scenario (ii) (Fig.~\ref{model})
which can explain both, the observational results of the molecular CO
being concentrated only along the {\em edges} of the kpc jet of
NGC\,4258, and also the {\em lack} of CO elsewhere in the galactic disk
of NGC\,4258: the molecular gas in galactic rotation interacts with the
jet's toroidal magnetic field due to ambipolar diffusion.
As stated above, we suggest that by this process the galactic molecular
gas becomes partly 'captured' by the magnetic field resulting in an
enhanced probability of finding the CO clouds along the jet ridges.
Part of the cloud material diffuses across the jet's toroidal magnetic field
and will finally hit the jet flow itself.
This matter will become shocked, the molecules will become dissociated,and the gas may be detected in H$\alpha$ across in the CO depleted funnel.
We note that also the apparent velocity change of the clumpy CO
components upstream and downstream of the jet (see Fig.~\ref{pv_minor})
suggests such an interaction.
As described in Sect.~\ref{sec:pv_minor} the velocity of the upstream
CO is about 50~${\rm km\,s^{-1}}$ lower than that of the downstream CO.

Such a velocity change is in both, its amount and direction with
respect to the galactic rotation, is comparable with the effect on rotating
gas in a trailing spiral density wave.
The streaming motions by density waves in the plane of the galaxy with respect
to purely circular motion has been found to be several km/s up to
$70 {\rm km\,s^{-1}}$ as observed for molecular clouds in
M~51 (Garc\'{\i}a-Burillo et al.\ 1993).
In NGC\,4258 we may observe similar effects on the CO velocities along both
edges of the inner {\em jet} and we see a concentration (a 'traffic jam') of CO
along the jets.
This scenario is especially feasible if the jet axis is slightly off
the galactic plane. Then the motion of the galactic molecular clouds
could only be influenced by the jet's toroidal magnetic field, which
extents to larger radii than the jet itself.

The detailed process of interaction is very difficult to investigate,
as a number of important parameters are not really known as the
magnetic field strength and orientation, the mass flow rate and
ionization fraction in the cloud.
Numerical magnetohydrodynamic simulations of the jet-cloud interaction
may finally prove or disprove our hypothesis.
A first check of our model suggestion of a jet-cloud interaction, however,
is to compare the energies involved in the jet and cloud motion as these
parameters should be in reasonable proportions.
The ratio of the {\em kinetic energy density} of a molecular cloud
to the jet magnetic field energy density is
\begin{equation}
\frac {E_{\rm kin, cloud}}{E_{B_{\rm jet}}} =
0.006\,n_{\rm cl}
\left(\frac{v_{\rm rot}}{200\,{\rm km\,s^{-1}}}\right)^2
\left(\frac{B_{\rm jet}}{200\,\mu{\rm G}}\right)^{-2}.
\end{equation}
where $n_{\rm cl}$ is the molecular hydrogen particle density.
The cloud on its orbital motion will compress the field until equilibrium
is reached.
For small densities ($n_{\rm cl}\leq 100\,{\rm cm^{-3}}$) the field energy
can exceed the cloud kinetic energy and may "brake" the cloud's motion.
A large molecular cloud may substantially deflect the jet motion,
an event which might have occurred in the northern end of the jet
(Plante et al.\ 1991).

Alignment of extragalactic jets with molecular gas is known for
other sources and sometimes connected to star formation
("jet-induced star formation").
Examples are Centaurus A (e.g. Mould et al.~2000,
Oosterloo \& Morganti 2005),
Minkowski's object (e.g. Croft et al.~2006 and references therein),
or highly redshifted quasars as BR~1202-0725 (Klamer et al.~2004).
Recent observation of the latter example have shown that all the
molecular gas in that source is concentrated in a compact
nuclear course (Riechers et al.~2006).
Jet-induced star formation is understood as mainly triggered by the
gas shocked along the propagating jet (Oosterloo \& Morganti 2005).

So far, these models do not explain the over-abundance of molecular
or atomic gas along the jets.
Our model of jet-ambient medium interaction by ambipolar diffusion may
explain the accumulation of molecular gas along the jets.
In the case of NGC~4258 we observe for the first time the accumulation
of CO gas along a jet in a nearby, spiral galaxy.
Star formation, however, has not (yet) been triggered along these jets.

\section{Summary and conclusions}
\label{sec:summa}

Using the IRAM interferometer at Plateau de Bure, we mapped the
\element[][12]{CO}(1--0) emission of the central part of NGC\,4258 along the
nuclear jet direction in the inner 3\,kpc.  We further observed NGC\,4258
in H$\alpha$ at the Hoher List Observatory of the University of Bonn
in order to ensure a proper relative positioning of the H$\alpha$ and
CO emission (Fig.~\ref{COonHalpha}).  The observational results can be
summarized as follows:

$-$ We detected two parallel CO ridges along a p.a.\  of $-24\degr$
with a total length of about $80\arcsec$ (2.8\,kpc)
(Fig.~\ref{mom0}). The ridges are separated by a CO depleted funnel
with a width of about $5\arcsec$ (175\,pc).

$-$ The H$\alpha$ emission is more extended and broader than the CO
emission with its maximum between the two CO ridges.  Contrary
to previous interpretations of Plante et al.\ (1991, 1994) we maintain
that the CO does
not outline 'walls' around the H$\alpha$ jets and hence does not
confine it. It rather seems to be mixed in location and in velocity
with the H$\alpha$ gas.

$-$ The CO iso-velocity map (Fig.\ref{mom1+0}) looks different
from that of a 'normal' spiral galaxy and indicates different
velocity components which show up more clearly in the p-v diagrams
along the galaxy's major axis (which is a good approximation for
the orientation along the CO jet features)(Fig.~\ref{pv_major}) as
parallel structures and in the pv-diagrams along the galaxy's minor
axis (Fig.~\ref{pv_minor}) as a blob structure. The main features
in the p-v diagrams along the galaxy's major axis (Fig.~\ref{pv_major})
rises, however, linearly with distance from the nucleus with red
shifted velocities in the northern half and blue shifted velocities
in the southern half as expected for CO gas in galactic rotation.

$-$ We further know from previous observations (Krause et al.\ 1990,
Cox \& Downes 1996) that the molecular gas in NGC\,4258 has only been
detected in the center and along the H$\alpha$ jets, at distances up
to 2\,kpc (Krause et al.\ 1990).  It has not been found in other parts of
the (central) galactic disk.

We propose a model in which the northern jet points {\em towards} the
observer (which also follows from the straightforward assumption that
the jet is perpendicular to the nuclear accretion disk). In this case
the CO emission is in normal but somewhat
modified galactic rotation. The
observed H$\alpha$ may be former molecular gas which has been shock
ionized during the partly violent interaction processes with the jets.

In our model the concentration of CO only along the
ridges is due to an {\em interaction} of the rotating gas clouds with
the jet's toroidal magnetic field (see Fig.~\ref{model}) which increases
the residence time of the molecular clouds near the jet when compared to
other loci in the galactic disk.  Both components, field and clouds,
are coupled by ambipolar diffusion.  We give some rough estimates of the
kinematic time scales for the jet itself and the jet-cloud interaction
on one hand and dynamical time scales of molecular clouds and the
ambipolar diffusion on the other.  The magnetic field strength
used for the estimates are inferred from observations near the nucleus
and further out up to kpc-scales together with theoretical jet models.
We conclude that the proposed interaction is quite feasible from a
theoretical point of view.

Concerning the observations, our model explains not only the
concentration of CO along the jets and correspondingly the lack of CO
elsewhere in the galactic disk. Furthermore it can explain the
relative positions of the blobs within a single p-v diagram
perpendicular to the galaxy's major axis and between the various p-v diagrams
from north to south.  The interaction of the molecular clouds with the
jet's toroidal magnetic field (i.e.\ magnetic interaction) is comparable
to the interaction of rotating gas with a spiral density wave
(i.e.\ gravitational interaction) which is responsible for the
appearance of the spiral arms in spiral galaxies.

In NGC~4258 we observed for the first time the accumulation of molecular gas along a jet in a nearby, spiral galaxy. A similar alignement of molecular gas along extragalactic jets is indicated in other more distant active galaxies and sometimes dicscussed in terms of "jet induced star formation". In NGC~4258, however, star formation has not (jet) been triggered along its jets.

\begin{acknowledgements}
It is a pleasure to thank J\"org Sanner and Tom Richtler from the
Sternwarte of the University of Bonn for the H$\alpha$ wide field
images which were essential for the detailed comparison of CO and
optical emission. We acknowledge  discussions with Dennis
Downes and Heino Falcke. We like to thank Phil Kronberg for fruitful discussions and helpful comments on the manuscript.
\end{acknowledgements}

\end{document}